\documentclass[aps,preprint]{revtex4}%
\usepackage{amsfonts}
\usepackage{amsmath}
\usepackage{amssymb}
\usepackage{graphicx}%
\begin{document}
\title{{\LARGE GENERAL RELATIVISTIC MACHIAN UNIVERSE }}
\author{Marcelo Samuel Berman$^{1}$}
\affiliation{$^{1}$Instituto Albert Einstein/Latinamerica\ - Av. Candido Hartmann, 575 -
\ \# 17}
\affiliation{80730-440 - Curitiba - PR - Brazil - email: msberman@institutoalberteinstein.org}
\date{(Original: 05 February, 2008. New Version: 1$^{st}$ October, 2008)}

\begin{abstract}
The Machian Universe, is usually described with Newtonian Physics, We give an
alternative General Relativistic picture for \ Mach's Universe. As such, we
show that, in the correct Machian limit, Schwarzschild's metric is coherent
with Robertson-Walker's, on condition that there be a cosmological constant,
or the Universe's rotation --- or both. It is now confirmed that the Universe
is accelerating, so the former condition applies. The latter was also
confirmed one more time with the recently discovered NASA space probes
anomalies. From Kerr-Lense-Thirring solution, we find an inverse scale-factor
dependent angular speed; we then, show that the cosmological "constant" may
have Classically\ \ originated from a centrifugal acceleration field.

\end{abstract}
\maketitle

\begin{center}

{\LARGE GENERAL RELATIVISTIC MACHIAN UNIVERSE}

\bigskip Marcelo Samuel Berman
\end{center}

\bigskip

{\large 1. INTRODUCTION }

\bigskip The purpose of this paper is (a) to review the Machian picture of the
Universe, according to a gravitational theory made with Special Relativity
plus Newtonian gravity; (b) to check a possible relation between
Schwarzschild's metric, and Robertson-Walker's, from the viewpoint of the
Machian picture in (a); (c) to analyze Lense-Thirring metric, and to interpret
the angular speed derived from it, in terms of the Machian picture above
(frame-dragging); (d) to state the equivalence between a cosmological (lambda)
"constant"\ field of accelerations, and the centrifugal one, originated by the
rotation of the Universe; (e) to conclude about the Machian limit of General
Relativity Theory, with a \ $R^{-1}$\ \ dependent angular speed, as also
calculated by this author in other prior papers.\ \ 

\bigskip

\bigskip The Machian intersection of General Relativity with Newtonian
Mechanics, is not necessarily in the low or weak gravitational limit.

We consider an extension for a paper by Barrow (Barrow, 1988): while he showed
that Newtonian Cosmology is equivalent with its General Relativistic version,
we shall show that Schwarzschild's metric and Robertson-Walker's, yield the
same results in the Machian limit. This is novel in the literature, and
extends our recently published papers (Berman, 2007b; 2008; 2008a,b), where we
have hinted on the rotation of the Universe. This last hypothesis could
explain the Pioneer anomaly, or, otherwise, the Pioneer anomaly could be the
experimental justification for the rotation of the Universe. Our explanation
of the Pioneer anomaly as a Machian effect of rotation, was intended to
explain, that an ubiquitous field of centripetal accelerations of cosmological
and Machian origin, existed for any observer in the Universe. A confirmation
of my theory was made by NASA, because not only the two Pioneer's 10 and 11
were subject to such anomaly, but also other space probes, except those in
closed orbits around the Sun. In fact, this shows that my explanation applies
well, \ and there is no contradiction with the inexistence of the anomaly in
closed orbits, because such motion is local, not cosmological. Remember that
hyperbolic orbits extend to infinity.

\bigskip

Berman (2008; 2008a,b) showed that Robertson-Walker's metric has hidden a
rotational state in addition to its expanding nature. As the cosmological
constant has now a relevant place in Modern Cosmology, both conditions
(non-null rotation and non-null cosmological constant) are possible. It will
be shown that, as well as Barrow's proof for the equivalence between Newtonian
and General Relativistic cosmologies, there is a kind of equivalence between
Schwarzschild's metric and Robertson-Walker's: the intersection represents the
Machian limit.

\bigskip

It will be studied, that the rotation of the Universe may respond for the
existence of a cosmological "constant", and that Kerr-Lense-Thirring metric
points out to an angular speed varying with the inverse of the scale-factor.

\bigskip

{\large \bigskip2. THE NEWTONIAN-MACHIAN PICTURE}

\bigskip

Barrow(1988) has shown that, with the adoption of conservation of energy,
according to Newtonian Mechanics, plus conservation of energy, according to
Thermodynamics, we obtain the usual Robertson-Walker's field equations for
energy density and cosmic pressure. The reason given is that: a) according to
General Relativity, local Physics means Newtonian gravitation; b)
Robertson-Walker's metric is homogeneous, so \ that, each point is equivalent
to any other, and thus, Newtonian gravitation applies everywhere.

\bigskip

If, then, Newtonian Physics means also cosmological Physics, we must take a
look, as we shall do below, on the possibility that Schwarzschild's metric
includes a limit for the validity of Robertson-Walker's metric. We shall
indeed show that for a Machian Universe, in the sense of Berman's
zero-total-energy approach (Berman, 2007; 2007a; 2007b; 2008; 2008b), this
requirement is fulfilled, so that, in fact, "local" means "global" phenomena.

\bigskip

The importance of the present paper resides in the demonstration that the
Machian-Newtonian view of the Universe, \ coincides in a certain sense, with
the Machian-General Relativistic picture, but of necessity, we have to
consider either the existence of a cosmological constant or, of a spin of the
Universe --- or both.

\bigskip

Consider a Newtonian description of a rotating sphere (the "causally connected
Universe"), endowed with a cosmological constant additional energy density.
The total energy is given by:

\bigskip

$E=E_{i}+E_{g}+E_{L}+E_{\Lambda}=0$ \ \ \ \ \ \ \ \ \ \ \ \ \ \ \ \ \ \ , \ \ \ \ \ \ \ \ \ \ \ \ \ \ \ \ \ \ \ \ \ \ \ \ \ \ \ \ \ \ \ \ \ \ \ \ \ \ (1)

\bigskip

where, the right hand side energy terms represent inertia \ ( $Mc^{2}$\ ),
\ \ gravitation \ \ \ \ \ \ ( $-\frac{GM^{2}}{R}$\ \ ) , rotation \ (
$\frac{L^{2}}{MR^{2}}$\ )\ \ \ and cosmological constant's \ ( $\frac{\Lambda
R^{3}}{6G}$\ )\ , where \ \ $L$\ \ , \ \ $\Lambda$\ \ , and \ \ $R$\ \ stand
respectively for the Universe's total spin, cosmological "constant", and its
"radius"\ . If we consider that relation (1) remains time-invariant ( $\dot
{E}=0$\ )\ , the only solution involving a time-varying scale factor \ (
$\dot{R}\neq0$\ )\ , implies the following generalised Brans-Dicke equalities:

\bigskip

$\frac{GM}{c^{2}R}=\gamma_{G}$
\ \ \ \ \ \ \ \ \ \ \ \ \ \ \ \ \ \ \ \ \ \ \ \ \ \ \ \ \ \ \ \ \ \ \ \ , \ \ \ \ \ \ \ \ \ \ \ \ \ \ \ \ \ \ \ \ \ \ \ \ \ \ \ \ \ \ \ \ \ \ \ \ \ \ \ \ \ \ \ \ \ \ \ \ \ (2)

\bigskip

$\frac{GL}{c^{3}R^{2}}=\gamma_{L}$%
\ \ \ \ \ \ \ \ \ \ \ \ \ \ \ \ \ \ \ \ \ \ \ \ \ \ \ \ \ \ \ \ \ \ \ \ \ , \ \ \ \ \ \ \ \ \ \ \ \ \ \ \ \ \ \ \ \ \ \ \ \ \ \ \ \ \ \ \ \ \ \ \ \ \ \ \ \ \ \ \ \ \ \ \ \ \ (3)

\bigskip

\bigskip and,

$\Lambda R^{2}=\gamma_{\Lambda}$%
\ \ \ \ \ \ \ \ \ \ \ \ \ \ \ \ \ \ \ \ \ \ \ \ \ \ \ \ \ \ \ \ \ \ \ \ \ , \ \ \ \ \ \ \ \ \ \ \ \ \ \ \ \ \ \ \ \ \ \ \ \ \ \ \ \ \ \ \ \ \ \ \ \ \ \ \ \ \ \ \ \ \ \ \ \ \ (4)

\bigskip

where all the \ $\gamma$\ 's\ \ are constants. This resembles Brans-Dicke
approximate Machian relation (Brans and Dicke, 1961). Of course, another
possibility would be \ $\dot{R}=0$\ \ , with another set of equalities.

\bigskip

In consequence, \ \ 

$\bigskip$

$R\propto M\propto L^{1/2}\propto\Lambda^{-1/2}$%
\ \ \ \ \ \ \ \ \ \ \ \ \ \ \ . \ \ \ \ \ \ \ \ \ \ \ \ \ \ \ \ \ \ \ \ \ \ \ \ \ \ \ \ \ \ \ \ \ \ \ \ \ \ \ \ \ \ \ \ \ \ \ \ \ \ \ (5)

\bigskip

The Machian picture, generally means that (2), (3), (4) and, thus, (5) apply
to the Universe, on condition that we remember the following:

\bigskip

\textbf{First}: (\textbf{sphericity postulate}) The Universe, for each and any
observer, resembles a "ball", of approximate spherical shape, of radius
\ $R$\ \ and mass \ \ $M$\ \ .

\bigskip

\textbf{Second: }(\textbf{egocentric observers' postulate}). \ Each observer
finds himself in the center of the "ball" . \ 

\bigskip

\textbf{Third}: (\textbf{observers' democratic principle}). As a consequence,
the Universe is homogenous so that, any location is equivalent to any other one.

\bigskip

\textbf{Fourth: (intersection of Newtonian gravity with Special Relativity).
}We have associated Einstein's relation from Special Relativity with Newtonian
gravity. General Relativity theory, would be similarly treated by Special
Relativity, but substituting Newtonian gravity, by other field equations.

\bigskip

{\large \bigskip3. GENERAL RELATIVISTIC-MACHIAN PICTURE}

\bigskip

Local Physics extends globally with the same "laws" -- if Newtonian Physics is
valid locally, it is also valid globally; but, we shall show that, if General
Relativity is applied locally through Schwarzschild's metric, it must be
equivalent to Robertson-Walker's in the large. This should be the General
Relativistic Machian picture.

\bigskip

Let us show that the above criterion is \ possible. Consider Schwarzschild's
metric with a cosmological term:

\bigskip

$ds^{2}=g_{00}dt^{2}-g_{00}^{-1}dr^{2}-d\sigma^{2}$
\ \ \ \ \ \ \ \ \ \ \ \ \ \ \ \ \ \ \ \ \ \ \ \ \ \ , \ \ \ \ \ \ \ \ \ \ \ \ \ \ \ \ \ \ \ \ \ \ \ \ \ \ \ \ \ \ \ \ \ (6)

\bigskip

\bigskip where,

$g_{00}=1-\frac{2GM}{c^{2}R}+\frac{GM\Lambda R}{3c^{2}}$
\ \ \ \ \ \ \ \ \ \ \ \ \ \ \ \ \ \ \ \ \ \ \ \ \ \ \ \ \ \ . \ \ \ \ \ \ \ \ \ \ \ \ \ \ \ \ \ \ \ \ \ \ \ \ \ \ \ \ \ \ \ \ \ \ \ \ \ \ \ \ (7)

\bigskip

If we associate the Schwarzschild's \ $g_{00}$\ with the Machian Brans-Dicke
equalities above, so that, any point of space is equivalent to any other, at
each one, we have a kind of Schwarzschild's \ \ $g_{00}$\ \ , given by:

\bigskip

$g_{00}=1-2\gamma_{G}+4\gamma_{\Lambda}=\Gamma>0$
\ \ \ \ \ \ \ \ \ \ \ \ \ \ \ \ \ \ , \ \ \ \ \ \ \ \ \ \ \ \ \ \ \ \ \ \ \ \ \ \ \ \ \ \ \ \ \ \ \ \ \ \ \ \ \ \ \ \ \ \ \ (8)

\bigskip

where \ \ $\Gamma$\ \ is a constant. A constant \ $g_{00}$\ , can be described
as Robertson-Walker's temporal coefficient (it can be always made equal to one).\ 

\bigskip

In order to give "power" to the above, we must understand that:

\bigskip

(a). the radial coordinate may be associated with the scale-factor in the
Machian limit.

(b). we can always reparametrize the metric coordinates by making:

\begin{center}
\bigskip

$dx^{\prime i}\equiv R^{2}(t)dx^{i}$ \ \ \ \ \ \ \ \ \ \ \ \ \ \ \ ,
\ \ \ \ \ \ \ \ \ \ \ \ ( $i=1,2,3$\ )

\end{center}

\bigskip and,

\bigskip

\ \ \ \ \ \ \ \ \ \ \ \ \ \ \ \ \ \ \ \ \ \ \ $dt^{\prime}\equiv dt$ \ \ \ \ \ \ \ \ \ \ \ \ \ \ \ \ \ \ \ \ \ \ \ \ \ \ \ .

\bigskip

\bigskip We now see why Robertson-Walker's metric (at least, the flat case) is
a kind of isomorphic to Schwarzschild's and, as we shall see bellow, its
rotating analogues, in the Machian limit.

\bigskip

\bigskip The reader can check the above procedure from Poisson's book
(Poisson, 2004).

We also see why Berman (2007c), has hinted that the Universe is to be
considered a white (or, generally speaking, black) hole, when General
Relativity with R.W.'s metric with cosmological constant, is taken into
account. The fact that there is a limit where \ both metrics yield the same
picture, has a similarity with Barrow's equivalence between Newtonian and
General Relativistic Cosmologies.

\bigskip

{\large 4. LENSE-THIRRING METRIC AND MACH'S PRINCIPLE}

\bigskip

\bigskip In order to match two spacetime metrics, each one valid in separate
but adjacent regions, there is a "thin-shell" formalism (Poison, 2004). So, we
first "freeze" the expansion of the Universe, keeping track only of the
rotation; the final results, in what refers to the rotation of the Universe,
are independent of the expansion, sufficing to remember that in Cosmology,
what we shall call as \ $R$\ \ bellow, is in fact an increasing function of time.

\bigskip

The "thin-shell" formalism (Poisson, 2004), begins with a comparison of
exterior and interior metrics for a rotating shell. In the low rotation limit,
the Kerr-metric gives origin Lense-Thirring description of the exterior of a
rotating shell:

\bigskip

$ds_{out}^{2}=-fdt^{2}+f^{-1}dr^{2}+r^{2}d$ $\Omega^{2}-\frac{4GMa}{r}\sin
^{2}\theta$ $d\phi$ $dt$ \ \ \ \ \ \ \ \ \ \ \ , \ \ \ \ \ \ \ \ \ \ \ \ \ \ \ \ \ \ \ \ \ (9)

\bigskip

where,

\bigskip

$f=1-\frac{2GM}{r}$ \ \ \ \ \ \ \ \ \ \ \ \ . \ \ \ \ \ \ \ \ \ \ \ \ \ \ \ \ \ \ \ \ \ \ \ \ \ \ \ \ \ \ \ \ \ \ \ \ \ \ \ \ \ \ \ \ \ \ \ \ \ \ \ \ \ \ \ \ \ \ \ \ \ \ \ \ \ \ \ \ \ \ \ \ \ (10)

\bigskip

\bigskip With decreasing values for \ $r$\ \ , we shall reach a "cut off"
radial value,

\bigskip

$r=R$ \ \ \ \ \ \ \ \ \ \ \ \ \ \ \ \ \ \ , \ \ \ \ \ \ \ \ \ \ \ \ \ \ \ \ \ \ \ \ \ \ \ \ \ \ \ \ \ \ \ \ \ \ \ \ \ \ \ \ \ \ \ \ \ \ \ \ \ \ \ \ \ \ \ \ \ \ \ \ \ \ \ \ \ \ \ \ \ \ \ \ \ \ \ \ \ \ (11)

\bigskip

\bigskip under which the shell is located, and where we find the "induced"
metric, \ as viewed from the exterior,

\bigskip

$ds_{\Sigma}^{2}=-Fdt^{2}+R^{2}d$ $\Omega^{2}-\frac{4GMa}{R}\sin^{2}\theta$
$d\phi$ $dt$ \ \ \ \ \ \ \ \ \ \ \ \ \ \ , \ \ \ \ \ \ \ \ \ \ \ \ \ \ \ \ \ \ \ \ \ \ \ \ \ \ \ (12)

\bigskip where,

$F=1-\frac{2GM}{R}$ \ \ \ \ \ \ \ \ \ \ \ \ . \ \ \ \ \ \ \ \ \ \ \ \ \ \ \ \ \ \ \ \ \ \ \ \ \ \ \ \ \ \ \ \ \ \ \ \ \ \ \ \ \ \ \ \ \ \ \ \ \ \ \ \ \ \ \ \ \ \ \ \ \ \ \ \ \ \ \ \ \ \ \ \ \ (13)

\bigskip The rotation parameter \ $a$\ \ \ is given in terms of the angular
momentum \ $J$\ \ \ and the mass \ $M$\ \ \ , by,

\bigskip

$a^{2}=\frac{J^{2}}{M^{2}}$\ \ \ \ \ \ \ \ \ \ \ \ \ \ \ \ \ . \ \ \ \ \ \ \ \ \ \ \ \ \ \ \ \ \ \ \ \ \ \ \ \ \ \ \ \ \ \ \ \ \ \ \ \ \ \ \ \ \ \ \ \ \ \ \ \ \ \ \ \ \ \ \ \ \ \ \ \ \ \ \ \ \ \ \ \ \ \ \ \ \ \ \ \ (14)

\bigskip

Consider now that we make the above metric, to include an opposite rotating
speed, by means of a diagonization, and described by,

\bigskip

$\psi=\phi-\omega_{0}t$ \ \ \ \ \ \ \ \ \ \ \ \ \ \ \ \ , \ \ \ \ \ \ \ \ \ \ \ \ \ \ \ \ \ \ \ \ \ \ \ \ \ \ \ \ \ \ \ \ \ \ \ \ \ \ \ \ \ \ \ \ \ \ \ \ \ \ \ \ \ \ \ \ \ \ \ \ \ \ \ \ \ \ \ \ \ \ \ (15)

\bigskip

so that, the diagonalization becomes,

\bigskip

$ds_{\Sigma}^{2}\approx-Fdt^{2}+R^{2}d$ $\bar{\Omega}^{2}$
\ \ \ \ \ \ \ \ \ \ \ \ \ \ , \ \ \ \ \ \ \ \ \ \ \ \ \ \ \ \ \ \ \ \ \ \ \ \ \ \ \ \ \ \ \ \ \ \ \ \ \ \ \ \ \ \ \ \ \ \ \ \ \ \ \ \ \ \ \ \ \ (16)

\bigskip

where, \bigskip\ 

$\bigskip$

$d$ $\bar{\Omega}^{2}\equiv R^{2}\left(  d\theta^{2}+\sin^{2}\theta d\psi
^{2}\right)  $ \ \ \ \ \ \ \ . \ \ \ \ \ \ \ \ \ \ \ \ \ \ \ \ \ \ \ \ \ \ \ \ \ \ \ \ \ \ \ \ \ \ \ \ \ \ \ \ \ \ \ \ \ \ \ \ \ \ \ \ \ \ \ \ \ (17)

\bigskip

The angular speed \ $\omega_{0}$\ \ , is found to be,

\bigskip

$\omega_{0}=\frac{2GM}{R^{3}}$ $a$\ \ \ \ \ \ \ \ \ \ \ \ \ \ \ . \ \ \ \ \ \ \ \ \ \ \ \ \ \ \ \ \ \ \ \ \ \ \ \ \ \ \ \ \ \ \ \ \ \ \ \ \ \ \ \ \ \ \ \ \ \ \ \ \ \ \ \ \ \ \ \ \ \ \ \ \ \ \ \ \ \ \ \ \ \ \ \ \ \ (18)

\bigskip

We now write the interior metric, which should be "cut-off" at the same radial
distance, which we now write as \ $\rho=R$\ \ \ \ ,

\bigskip

$ds_{int}^{2}\approx-Fdt^{2}+d\rho^{2}+\rho^{2}d$ $\bar{\Omega}^{2}$
\ \ \ \ \ \ \ \ \ \ \ \ \ \ . \ \ \ \ \ \ \ \ \ \ \ \ \ \ \ \ \ \ \ \ \ \ \ \ \ \ \ \ \ \ \ \ \ \ \ \ \ \ \ \ \ \ \ \ \ \ \ \ \ \ (19)

\bigskip

By working with a perfect fluid model, Poisson shows that the shell must move
with angular speed \ $\omega$\ \ , \ given by,

\bigskip

\bigskip$\omega=\omega_{0}\left[  \frac{1-F}{\left(  1-\sqrt{F}\right)
\left(  1+3\sqrt{F}\right)  }\right]  $ \ \ \ \ \ \ \ \ \ \ \ \ \ \ \ \ \ . \ \ \ \ \ \ \ \ \ \ \ \ \ \ \ \ \ \ \ \ \ \ \ \ \ \ \ \ \ \ \ \ \ \ \ \ \ \ \ \ \ \ \ \ \ \ \ \ \ \ \ \ (20)

\bigskip

\bigskip We now consider an observer at constant \ $\psi$\ \ \ , \ moving
then, with angular speed,

\bigskip

$\omega_{0}=\frac{d\phi}{dt}$ \ \ \ \ \ \ \ \ \ . \ \ \ \ \ \ \ \ \ \ \ \ \ \ \ \ \ \ \ \ \ \ \ \ \ \ \ \ \ \ \ \ \ \ \ \ \ \ \ \ \ \ \ \ \ \ \ \ \ \ \ \ \ \ \ \ \ \ \ \ \ \ \ \ \ \ \ \ \ \ \ \ \ \ \ \ \ \ \ \ \ \ \ \ \ \ (21)

\bigskip

This rotation is relative to the "fixed" stars, obtained from the exterior
metric, which is Lorentzian at infinity, i.e., when \ $r\longrightarrow\infty
$\ \ , we retrieve Schwarzschild's metric. In the non-rotating frame, the
shell's angular speed is given by,

\bigskip

$\omega_{1}=\frac{d\phi}{dt}=\frac{d\psi}{dt}+\omega_{0}=\omega_{0}\left[
\frac{1+2\sqrt{F}}{\left(  1-\sqrt{F}\right)  \left(  1+3\sqrt{F}\right)
}\right]  $ \ \ \ \ \ \ \ \ \ \ \ \ .\ \ \ \ \ \ \ \ \ \ \ \ \ \ \ \ \ \ \ \ \ \ \ \ \ \ \ \ \ \ \ \ \ \ \ \ \ \ (22)\ 

\bigskip

In the Machian Universe, it is the fixed stars that are no more fixed, i.e.,
they rotate relative to the shell, and from relations (2), (3), (4) and (5),
\ we have,

$\bigskip$

$F=1-2\gamma_{G}=$\ \ constant\ \ \ \ \ \ \ \ , \ \ \ \ \ \ \ \ \ \ \ \ \ \ \ \ \ \ \ \ \ \ \ \ \ \ \ \ \ \ \ \ \ \ \ \ \ \ \ \ \ \ \ \ \ \ \ \ \ \ \ \ \ \ \ \ \ \ \ \ \ \ \ \ \ \ \ \ (23)

\bigskip

and, then,

\bigskip

$\omega_{1}=\alpha$ $\omega_{0}$ \ \ \ \ \ \ ( \ $\alpha$\ \ = constant )
\ \ \ \ \ . \ \ \ \ \ \ \ \ \ \ \ \ \ \ \ \ \ \ \ \ \ \ \ \ \ \ \ \ \ \ \ \ \ \ \ \ \ \ \ \ \ \ \ \ \ \ \ \ \ \ \ \ \ \ \ \ \ \ (24)

\bigskip

For \ $\gamma_{G}\cong\frac{1}{2}$\ \ \ , \ \ we have \ \ $F\cong0$\ \ \ \ ,
and, then,

\bigskip

$\omega_{1}\cong\omega_{0}=\frac{2GM}{R^{3}}$ $a$%
\ \ \ \ \ \ \ \ \ \ \ \ \ \ \ , \ \ \ \ \ \ \ \ \ \ \ \ \ \ \ \ \ \ \ \ \ \ \ \ \ \ \ \ \ \ \ \ \ \ \ \ \ \ \ \ \ \ \ \ \ \ \ \ \ \ \ \ \ \ \ \ \ \ \ \ \ \ \ \ \ \ \ \ \ \ (25)

\bigskip

or,

\bigskip

$J=a$ $M=\frac{\omega_{1}}{2}$ $R^{3}$ \ \ \ \ \ \ \ \ \ \ \ \ \ \ . \ \ \ \ \ \ \ \ \ \ \ \ \ \ \ \ \ \ \ \ \ \ \ \ \ \ \ \ \ \ \ \ \ \ \ \ \ \ \ \ \ \ \ \ \ \ \ \ \ \ \ \ \ \ \ \ \ \ \ \ \ \ \ \ \ \ \ \ \ \ \ (26)

\bigskip

The Machian spin of the Universe \ \ $L$\ \ , was obtained as being
proportional to \ $R^{2}$\ \ , so that, the Machian angular speed, is
proportional to \ $R^{-1}$\ \ , according to (26) .

\bigskip

The \ $R^{-1}$\ \ dependence of the angular speed, has been found by Berman in
several other contexts, (Berman, 2007b; 2008a; 2008b; 2008c; 2008d).

\bigskip

{\large \bigskip5. CLASSICAL ORIGIN FOR THE DARK ENERGY}

\bigskip In last Section, we must remember to say that the condition for the
signature of the metric not to be altered, is that \ \ $F>0$\ \ \ , or
\ \ \ $\gamma_{G}<\frac{1}{2}$\ \ \ . \ 

\bigskip

From Raychaudhuri's equation for a perfect fluid, we obtain the following
Robertson-Walker's metric result,

\bigskip

$\ddot{R}=\left[  -\frac{\kappa}{6}\left(  \rho+3p\right)  +\frac{1}{3}%
\Lambda\right]  $ $R$ \ \ \ \ \ \ \ \ \ \ \ \ \ \ \ . \ \ \ \ \ \ \ \ \ \ \ \ \ \ \ \ \ \ \ \ \ \ \ \ \ \ \ \ \ \ \ \ \ \ \ \ \ \ \ \ \ \ \ \ \ \ \ \ \ \ \ \ \ \ \ (27)

\bigskip

It is clear that the cosmological term represents the repulsive acceleration,

\bigskip

$a_{\Lambda}=\frac{1}{3}\Lambda$ $R$ \ \ \ \ \ \ \ \ \ \ . \ \ \ \ \ \ \ \ \ \ \ \ \ \ \ \ \ \ \ \ \ \ \ \ \ \ \ \ \ \ \ \ \ \ \ \ \ \ \ \ \ \ \ \ \ \ \ \ \ \ \ \ \ \ \ \ \ \ \ \ \ \ \ \ \ \ \ \ \ \ \ \ \ \ \ \ \ \ \ \ \ \ \ (28)

\bigskip

On the other hand, we have shown that the Universe undergoes a rotational
state, so that, there is a Machian centrifugal acceleration,

\bigskip

$a_{cf}=\omega^{2}R$ \ \ \ \ \ \ \ \ \ \ \ \ \ . \ \ \ \ \ \ \ \ \ \ \ \ \ \ \ \ \ \ \ \ \ \ \ \ \ \ \ \ \ \ \ \ \ \ \ \ \ \ \ \ \ \ \ \ \ \ \ \ \ \ \ \ \ \ \ \ \ \ \ \ \ \ \ \ \ \ \ \ \ \ \ \ \ \ \ \ \ \ \ \ \ \ (29)

\bigskip

From the Machian relations (2), (3), (4) and (5), and, from the result of last
Section, \ that the angular speed depends on \ $R^{-1}$\ , we find that both
accelerations above, depend on \ $R^{-1}$\ , and are, from the Machian
viewpoint, equivalent. Do not forget that \ \ $\Lambda$\ \ \ and \ $\omega
^{2}$\ \ should depend on \ \ $R^{-2}$\ \ .\ \ \bigskip

It does not matter if we say that the Universe rotates, or that the Universe
is endowed with a cosmological "constant" term. We would say that the spin of
the Universe stands for the Classical origin of the cosmological constant; and
we need not refer to Quantum phenomena in order to generate a lambda-Universe.\ \ 

\bigskip

{\large \bigskip6. CONCLUSIONS AND PREDICTIONS}

The main result of this paper, depends directly on relation (8); this relation
implies that local \ \ $g_{00}$\ \ is equal to a positive constant, when
applied for the Universe, and this entails either the rotation of the Universe
or a given cosmological constant, or both. The latter, implies a deep new look
on cosmological theories, and in Berman (2008; 2008a,b),\ he\ showed that
Robertson-Walker's metric has a hidden rotational character, along with the
usual evolutionary property. \bigskip The lambda-accelerating Universe was
also confirmed by recent observations.

\bigskip

That being the case, we predict that the left-handed Universe, is caused by
rotation; global statistical analysis of rotating clusters of galaxies, and
chaotic motions in the Universe, must show the left-handed property, due to
rotation. All phenomena, like violation of parity and matter-antimatter
asymmetry, are explained likewise. A lambda Universe means perhaps a rotating
one. This makes us predict that, not only the Universe is left-handed (Barrow
and Silk, 1983), but if it will be paid attention, to chaotic phenomena in the
Universe, and, also to rotational states of galaxies and clusters of them, a
preference for the left-hand must be found. Not only, \ parity violations, but
also barion-antibarion asymmetries (Feynman et al., 1962) will be explained in
terms of the said rotation of the Universe.

\bigskip

{\large Acknowledgements}

\bigskip

The author thanks, an anonymous referee, for important contributions, his
intellectual mentors, Fernando de Mello Gomide and the late M. M. Som, and
also to Marcelo Fermann Guimar\~{a}es, Nelson Suga, Mauro Tonasse, Antonio F.
da F. Teixeira, and for the encouragement by Albert, Paula and Geni.

\bigskip

\bigskip{\Large References}

\bigskip Barrow, J.D. (1988) - \textit{The Inflationary Universe, }in
\textit{Interactions and Structures in Nuclei, }pp. 135-150, eds. R. Blin
Stoyle \& W. D. Hamilton, Adam Hilger, Bristol.

Barrow, J.D; Silk, J. (1983) - \textit{The Left Hand of Creation: The Origin
and Evolution of Expanding Universe, }Basic Books, New York.

\bigskip Berman,M.S. (2007) - \textit{Introduction to General Relativistic and
Scalar Tensor Cosmologies}, Nova Science, New York.

Berman,M.S. (2007a) - \textit{Introduction to General Relativity and the
Cosmological Constant Problem}, Nova Science, New York.

Berman,M.S. (2007b) - \textit{The Pioneer Anomaly and a Machian Universe,
}Astrophysics and Space Science, \textbf{312}, 275.

Berman,M.S. (2007c) - \textit{Is the Universe a White-Hole?}, Astrophysics and
Space Science, \textbf{311}, 359.

\bigskip Berman,M.S. (2008) - \textit{Shear and Vorticity in a Combined
Einstein-Cartan-Brans-Dicke Inflationary Lambda-Universe, }Astrophysics and
Space Science, \textbf{314,} 79-82. For a previous version, see Los Alamos
Archives - \ http://arxiv.org/abs/physics/0607005

\bigskip Berman,M.S. (2008a) -\textit{ A General Relativistic Rotating
Evolutionary Universe - Part II, }Astrophysics and Space Science, \textbf{315,
}367-369. Posted with another title, see Los Alamos Archives -
http://arxiv.org/abs/0801.1954 [physics.gen-ph]\ 

Berman,M.S. (2008b) -\textit{ A General Relativistic Rotating Evolutionary
Universe, }Astrophysics and Space Science, \textbf{314}, 319-321. For an
earlier version, see Los Alamos Archives - http://arxiv.org/abs/0712.0821\ 

\bigskip Berman,M.S. (2008c) -\textit{ A Primer in Black Holes, Mach's
Principle and Gravitational Energy, }Nova Science Publishers, New York.

Berman,M.S. (2008d) -\textit{ On the Machian Origin of Inertia, }Astrophysics
and Space Science, \textbf{318, }269-272.

\bigskip Brans, C.; Dicke, R.H. (1961) - Physical Review, \textbf{124}, 925.

Feynman, R.P. et al (1962) - \textit{Lectures on Physics}, Addison-Wesley, Reading.

Poisson, E. (2004) - \textit{A Relativistic Toolkit, }CUP, Cambridge (p. 85; 94-95).

\end{document}